\documentclass[11pt]{article}

\usepackage{sectsty}
\usepackage{graphicx}
\PassOptionsToPackage{hyphens}{url}\usepackage[colorlinks=true,linkcolor=blue,anchorcolor=blue,citecolor=blue,filecolor=blue,menucolor=blue,runcolor=blue,urlcolor=blue]{hyperref}
\usepackage[hyphenbreaks]{breakurl}
\usepackage[skip=10pt plus1pt]{parskip}
\usepackage{titlesec}
\usepackage{booktabs}
\usepackage{scrextend}
\usepackage{setspace}
\usepackage[font=tiny,labelfont=bf]{caption}

\newcommand{\quotetext}[1]{
    \bigskip
    \begin{addmargin}[4em]{4em}
    \doublespacing
        {\fontsize{22bp}{35}\selectfont {``#1''}}
    \end{addmargin}
}
\newcommand{\quoteattr}[1]{
    \begin{addmargin}[4em]{4em}
    {\fontsize{12pt}{35}\selectfont #1}
    \end{addmargin}
    \bigskip
}
\title{\huge{A framework for improving the accessibility \\of research papers on arXiv.org}}
\author{ Shamsi Brinn, Christopher Cameron, David Fielding, Charles Frankston, \\Alison Fromme, Peter Huang, Mark Nazzaro, Stephanie Orphan, Steinn Sigurdsson, \\Ryan Tay, Miranda Yang, Qianyu Zhou }
\date{\today}

\begin{document}
\maketitle
\begin{addmargin}[4em]{4em}
\begin{abstract}
\doublespacing
The research content hosted by arXiv is not fully accessible to everyone due to disabilities and other barriers. This matters because a significant proportion of people have reading and visual disabilities, it is important to our community that arXiv is as open as possible, and if science is to advance, we need wide and diverse participation. In addition, we have mandates to become accessible, and accessible content benefits everyone. In this paper, we will describe the accessibility problems with research, review current mitigations (and explain why they aren’t sufficient), and share the results of our user research with scientists and accessibility experts. Finally, we will present arXiv’s proposed next step towards more open science: offering HTML alongside existing PDF and TeX formats. An \href{https://info.arxiv.org/about/accessibility_research_report.html}{accessible HTML version of this paper} is available on the arXiv website.
\end{abstract}
\end{addmargin}

\pagebreak

\section*{Introduction}
Improving access to research is a broad and inclusive effort, championed and moved forward by individuals and organizations around the world. In scientific publishing, arXiv has played an important role in open access for over 30 years by removing financial, institutional, and geographic barriers to research.

Truly open access, though, means more than free and available. When we say that arXiv is open we also must ask: open to whom? Accessibility is the practice of \emph{ensuring access regardless of disability}. It is the next frontier in the open science movement.

\section{Barriers to access}
Barriers to accessing research are broad and can be related to a permanent or temporary disability, or situational factors. Vision impairments and learning disabilities can impede access to written material. If content display is not flexible then using a mobile device can be a barrier. Language barriers or lack of internet access are broadly experienced impediments to accessing research.

Countless people around the globe encounter such barriers. People with disabilities are the world’s largest minority (United Nations). More than a quarter of the world’s population has a diagnosed vision impairment (World Health Organization), and vision loss is among the top ten disabilities in the United States (CDC). For much of the world, including english speaking deaf, written English is not their first language (Hrastinski and Wilbur). 20\% of people in the United States have dyslexia (Yale Center for Dyslexia and Creativity), and 26\% of people in the United States self-report living with at least one disability (Elavsky et al.). A wide array of temporary and situational impairments (such as bright or low light conditions, or a temporary loss of vision) affect many more who do not have a permanent disability diagnosis.

\textbf{“Is there really open and immediate access for everyone, if scholars and students with disabilities cannot access and use research articles?”}(Wentz et al.) 

In the United States, 14.5\% of K-12 students have a recognized disability (National Center for Education Statistics). Ever increasing numbers of high school students are participating in Computer Science curriculums (Vegas and Fowler) and other STEM fields, so naturally the percent of students with disabilities in these courses is proportionally increasing as well (Code.org). These efforts will, over time, lead to more students with disabilities in higher education who will fully expect and demand equal access to the research output in their field.

Furthermore, in the United States there are clear mandates for equivalent access to websites and federally funded research. The need to make research more accessible to all is evident and pressing.

\quotetext{We can only do so much to welcome people, but if they can’t participate fully it's all just nice talk.}
\quoteattr{—Dr. Kimberly Arcand, data visualizer and science communicator}

Feedback from arXiv users confirms the barriers to current participation:

\textbf{“There are not that many blind mathematicians and scientists because of all the stumbling blocks.”}

\textbf{“Most people veer away from STEM subjects towards text subjects because until a lot more content is accessible they need to work really hard to find it.”}

Despite the wealth of data around needs, and a web of regulations mandating equal access, the vast majority of research papers have low levels of accessibility, creating significant barriers for a large number of people.

\section{Mandates}
Access is a critical effort in academia. In the USA, the White House recently released a policy \href{https://www.whitehouse.gov/ostp/news-updates/2022/08/25/ostp-issues-guidance-to-make-federally-funded-research-freely-available-without-delay/}{mandating open access for all federally funded research}, while \href{https://www.coalition-s.org/}{Plan S} in Europe aims for no science to be locked behind paywalls. Internationally, we have the \href{https://www.go-fair.org/fair-principles/}{FAIR principles} (of which arXiv is a signatory), and the \href{https://en.unesco.org/science-sustainable-future/open-science/recommendation}{UNESCO recommendations on open science}, among others. The movement towards open access is making clear and steady progress.

The movement towards accessibility of research, however, has stalled. Accessibility should be part of every conversation about access. Accessibility is a requirement for most federally funded research in many countries, but in practice only a very low percentage of research papers pass accessibility criteria (Wang et al.). And of these, even fewer pass a more careful human review (Elavsky et al.). 

There are formidable obstacles for researchers with disabilities to enter and participate in STEM fields. Foremost among them is equal access to research and data.

\quotetext{The biggest barrier for me is the time you need to spend tracking down formats you can access in order to gain the same information available to everybody else.}
\quoteattr{—Robin Williams, statistician}

Existing legal mandates include making websites and federally funded content accessible to all users. Further regulations continue to gain momentum at the Federal level in the USA (ADA Title III). Furthermore, educational experts have been hard at work for decades transforming primary education to be more inclusive, which will put pressure on all areas of higher education as more diverse students enter STEM, including on repositories of scientific research like arXiv.
\pagebreak

\quotetext{There is work going on earlier in the pipeline, for example making mathematical language accessible to high schoolers. As more people with disabilities enter STEM fields, students will have the expectation of accessing information.}
\quoteattr{—Joe Zesski, Assistant Director, Northeast ADA Center}

Most importantly, arXiv has a mandate from our global community to improve access. Scientists need this and the research community is asking for it. One way of meeting these mandates is through well formatted HTML.

HTML is also the foundation of machine readability with major repercussions for the future of scientific discovery. The science that researchers share on arXiv is important, and automated research flows using machine learning are increasingly relied upon for accelerating discovery.(Policy and Global Affairs et al.) Semantic markup and accessibility for assistive technology is a good first step towards ensuring the arXiv repository better supports emerging approaches to conducting science.

arXiv’s long-term mission is simply to serve the needs of the scientific community. Everyone should be able to participate in the wealth of scientific knowledge contributed to arXiv by researchers from all over the world. Accessibility is the next logical step.

\quotetext{Research should be accessible for everyone in the broadest possible way.}
\quoteattr{—Dr. Shiri Azenkot, Associate Professor and Accessibility researcher}

\section{Current state of research accessibility}
Scientific research is primarily shared in PDF format, which mimics the printed page. It has been the primary format for decades, so any analysis of the accessibility of research inevitably becomes—in part—an analysis of the PDF format itself.

For authors, PDF offers a reliable visual format for sharing and disseminating their work. For publishers, it offers an archival format that won’t change over time. The accessibility of PDFs can be improved by tagging and adding descriptions.

However, the format has serious limitations that have had a profound impact on the accessibility of research. PDF is not a suitable format for the web and has low native accessibility. PDFs are challenging for people with a variety of reading disabilities, including blindness, low vision, dyslexia, and more. It is time consuming to improve the accessibility of PDFs, and even if this work is done it has no effect on its poor mobile performance.

Research from the Allen AI institute reports a miserably low accessibility rate for research papers:

\textbf{“PDF accessibility adherence is low across all fields of study. Of the five accessibility criteria we assess, only 2.4\% of the PDFs we assess demonstrate full compliance.”}(Wang et al.)

\quotetext{The best PDF will ever achieve is what HTML delivers. All it can do is catch up.}
\quoteattr{—Dr. Jonathan Godfrey, Senior Lecturer in Statistics}

Screen readers rely on semantic markup (for example headers, images, formulas, and so on) to correctly interpret content. PDFs do not natively include semantic properties, and it must be tagged with this information after the fact to make it accessible. Tagging is time consuming, takes specialized knowledge, and requires proprietary tools. These tools (and the expertise to use them) are not free, nor are they intuitive. Adobe’s own manual for making PDFs accessible is 94 pages long! (Elavsky et al.) In addition, if you find you need to regenerate the PDF again for any reason, you must tag the document again from scratch.

It is perhaps not surprising then that efforts to promote tagging PDFs for accessibility have not become the norm in academic publishing.

\textbf{“In CHI 2014, a year in which considerable effort was spent giving author’s feedback on the accessibility of their documents, only 26.8\% included any document tags at all.”}(Bingham et al.)

\textbf{“The criterion with the lowest rate of compliance is Alt-text, which has remained stable between 5–10\% and has been lower in recent years. Since Alt-text is the only criterion of the five which always necessitates author intervention, we believe this is a sign that authors have not become more attuned to accessibility needs.”}(Wang et al.)

Even when this manual work is done the resulting PDF is still only partially accessible: Two column layouts often confuse screen reader software; Text and graphics don’t reflow in mobile devices or with magnification; There is limited parsability for third party tools; And if math and data visualizations did have rich markup in the original source—such as a TeX formula or a SVG graph—the data are often flattened when the PDF is generated.

The typesetting software used to generate PDF results in differing levels of accessibility, with Microsoft Word producing the highest level of compliance, and LaTeX producing the lowest (Wang et al.). In the scientific fields that arXiv covers, LaTeX is widely used, and 90\% of arXiv’s corpus in recent years was submitted as TeX source. The LaTeX core team is working on generating more accessible PDFs, and we plan to incorporate this work into arXiv when it is released.

On mobile devices, PDFs are far behind standard and don’t meet basic user expectations. Research done by Adobe found that 65\% of Americans find consuming content on mobile frustrating; 45\% stopped reading or didn’t even try; and 72\% say they would work on their mobile devices more if it were easier to read documents (Adobe). Mobile needs in academia vary widely, from a researcher catching up on a paper while traveling, to one with limited access to technology using their mobile phone as their only option for reading papers. Flexible reflow of content is a must to expand access and efficiency.

ACM conference have also found low accessibility rates of PDF, despite significant effort and have chosen HTML5 format as a goal to \textbf{“ultimately make accessibility easier and more standardized.”} (Mankoff, et al.)

\quotetext{You can make PDF accessible, but screen readers are much more efficient when working with HTML. To tag PDF you need specialized skills and tools. For HTML, all these things are comparatively easier.}
\quoteattr{—Avneesh Singh, chair, accessibility task force - W3C EPUB 3 Working Group and Publishing Community Group}

PDF is the current standard, but our user research tells us that providing HTML—in addition to PDF and TeX source—will substantially increase accessibility to research posted to arXiv. Well formatted HTML will support and empower the many different ways that scientists consume research data.

\subsection{Existing tools}
A great deal of work has been done by sung and unsung heroes in the accessibility space, and as arXiv explores our role in making research more accessible we know that we are standing on the shoulders of giants. Critical and transformative work has been achieved in the standardizing and formatting of math on the web; converting TeX to HTML (and PDF, Word and other formats to HTML as well); guiding researchers in writing alternative content for figures and images; making data in charts and graphs parseable by screen readers; translation services (including for Braille); identifying and measuring accessibility on the web; and the incredibly important and advanced work that has gone into screen readers and haptic displays.

We also want to take the opportunity to acknowledge the countless scientists with disabilities who have invented and contributed to tools, standards, and community understanding over their careers. We have spoken to so many scientists who had to first invent or build the tools they use just to participate in their field. They have paved the way for countless others, as well as for the assistive software in use today. They donate their time to open source tools, guide new scientists in navigating the confusing landscape of access, generously share their expertise (including with arXiv), and endlessly advocate for equal access. All the while overcoming the daily obstacles they face in their own work. Thank you.

To the professors and teachers helping students overcome the formidable barriers built into how we share and publish research, thank you. To the researchers who make their work accessible, thank you.

We can make things easier and more equal for our colleagues. Making research available to everyone regardless of disability is the next stage of Open Science.

\subsection{The tool we need}
You might ask, with so many tools already built, why do we need anything else? After thorough analysis we found that none of the existing tools, as they are, can provide a smooth experience for authors and readers. Authors should be able to submit their work with the software they currently use and without being an accessibility expert. To achieve that will require new technological solutions as well as cultural changes beyond arXiv, but we have an important role to play.

Our goal is to be able to say to the arXiv community: you bring your expertise in your field; we will help close the gap on all the rest. Thanks to the work of so many in the broad and complex field of accessibility, we believe this is achievable today, though still not easy.

\section{Researching accessibility in the arXiv ecosystem}
We undertook to research the experience of people with a variety of disabilities and other barriers as it relates to accessing research articles, on arXiv and beyond. We also gathered input from experts in different fields: accessibility researchers; writers of web standards for accessibility, Math, and PDF; TeX and LaTeX experts; developers of screen reader and other assistive technologies; scholars of accessibility law; and science communicators.

Our research took two forms, a quantitative survey and a series of qualitative interviews.

\subsection{Survey}
We developed a survey to investigate behavior and preferences related to accessing content on arXiv.

\subsubsection{Process}
We invited two groups of people to take this survey: researchers who directly rely on assistive technology, and professors and colleagues who assist researchers in accessing the research and data they need.

Out of a pool of 275 volunteers we had a response rate of 18\% and received a total of 53 individual responses. Though this number appears small, it is above the industry expected ~6\% when taking into account the long length and detail of the survey. The arXiv community again proved its generosity of spirit.

\subsubsection{Demographics}
Our respondents were primarily frequent arXiv users. 58\% use arXiv daily and 25\% weekly. 25\% of respondents directly use assistive technology, while 75\% are not direct users, but are involved in assisting people who are.

Respondents represent a variety of fields, with the highest number of respondents (*42\%) from Physics, followed by Math and Computer Science.

\begin{table}[!ht]
    \centering
    \begin{tabular}{ll}
    \hline
        Field
    & Percentage
    \\ \hline
        Physics
    & 31\%
    \\ \hline
        Mathematics
    & 21\%
    \\ \hline
        Computer Science
    & 13\%
    \\ \hline
        Other
    & 13\%
    \\ \hline
        Astrophysics
    & 12\%
    \\ \hline
        Engineering
    & 6\%
    \\ \hline
        Biology
    & 4\%
    \\ \hline
    \end{tabular}
\end{table}

Respondents come from various, but not all, geographic regions. Europe had the highest representation at 44\%, followed by Asia at 23\%, North America at 21\%, South America at 8\%, and Africa at 4\%. We continue to work towards greater global representation in our surveys.

The assistive technology reported in use by our respondents are:

\begin{table}[!ht]
    \centering
    \begin{tabular}{ll}
    \hline
        Assistive technology
    & Percentage
    \\ \hline
        Screen reader
    & 43\%
    \\ \hline
        Adjust screen color/contrast
    & 24\%
    \\ \hline
        Magnifier
    & 24\%
    \\ \hline
        Voice command
    & 10\%
    \\ \hline
    \end{tabular}
\end{table}

\subsubsection{Survey Results}

\textsc{Access to research output}

\textbf{Summary:} Our respondents heavily depend on access to research, but users of assistive technology report they only have access to 38\% of the research they need without assistance. Overall, participants reported that access has improved somewhat over the last five years.

Our respondents heavily depend on access to research, with 89\% saying that research is completely or somewhat essential to their professional work:

\begin{table}[!ht]
    \centering
    \begin{tabular}{lll}
    \hline
        Dependence level
    & Yes, use assistive technology
    & No, don’t use assistive technology
    \\ \hline
        Completely Essential
    & 80\%
    & 78\%
    \\ \hline
        Somewhat Essential
    & 10\%
    & 11\%
    \\ \hline
        Somewhat Optional
    & 10\%
    & 8\%
    \\ \hline
        Completely Optional
    & 0\%
    & 2\%
    \\ \hline
    \end{tabular}
\end{table}

We next asked what level of access respondents have today, without requiring assistance from others. Overall the numbers were high: 89\% report having access to all or most of the research they need without assistance. However, those numbers look quite different if the respondent uses assistive technology, with only 30\% reporting access to all papers without assistance:

\begin{table}[!ht]
    \centering
    \begin{tabular}{lll}
    \hline
        Current access
    & Yes, use assistive technology
    & No, don’t use assistive technology
    \\ \hline
        All research is accessible
    & 30\%
    & 56\%
    \\ \hline
        Most research is accessible
    & 40\%
    & 38\%
    \\ \hline
        About half is accessible
    & 10\%
    & 6\%
    \\ \hline
        Most is not accessible
    & 20\%
    & 
    \\ \hline
    \end{tabular}
\end{table}

We also asked whether access to research had improved in the last five years, and for the majority of users, including those who use assistive technology, it has improved at least somewhat:

\begin{table}[!ht]
    \centering
    \begin{tabular}{lll}
    \hline
        Improvement level
    & Yes, use assistive technology
    & No, don’t use assistive technology
    \\ \hline
        It has improved a lot
    & 40\%
    & 47\%
    \\ \hline
        It has improved a little
    & 20\%
    & 24\%
    \\ \hline
        It is about the same
    & 30\%
    & 26\%
    \\ \hline
        It is worse
    & 10\%
    & 3\%
    \\ \hline
    \end{tabular}
\end{table}

\rule{480pt}{.5pt}
\textsc{Barriers to access}

\textbf{Summary:} Survey respondents agree that PDF formatting is the biggest barrier. The main reason reported for not submitting accessible papers is a lack of understanding around requirements.

When asked what the biggest barriers were to accessing papers, PDF formatting topped the list:

\begin{table}[!ht]
    \centering
    \begin{tabular}{lll}
    \hline
        Barriers
    & Yes, use assistive technology
    & No, don’t use assistive technology
    \\ \hline
        PDF formatting
    & 22\%
    & 25\%
    \\ \hline
        Images
    & 15\%
    & 10\%
    \\ \hline
        Math
    & 12\%
    & 12\%
    \\ \hline
        Graphs and charts
    & 10\%
    & 11\%
    \\ \hline
        TeX macros
    & 7\%
    & 10\%
    \\ \hline
        Language
    & 7\%
    & 5\%
    \\ \hline
        Font size
    & 7\%
    & 6\%
    \\ \hline
        Colors
    & 7\%
    & 2\%
    \\ \hline
        Other
    & 5\%
    & 4\%
    \\ \hline
        Contrast
    & 5\%
    & 4\%
    \\ \hline
        arxiv.org website
    & 3\%
    & 11\%
    \\ \hline
    \end{tabular}
\end{table}

The arxiv.org website is not a significant barrier to users of assistive technology, which our interviews also highlighted for us and is good news. But we do not ignore that the website has barriers unrelated to accessibility (these include limited coverage of related fields and a poor search function), though beyond the scope of this report.

We asked respondents the reasons that stopped them from submitting accessible papers.
\pagebreak
\begin{table}[!ht]
    \centering
    \begin{tabular}{lll}
    \hline
        Reason
    & Yes, use assistive technology
    & No, don’t use assistive technology
    \\ \hline
        Their own papers are \\
        accessible already
    & 50\%
    & 21\%
    \\ \hline
        Not knowing what an \\
        accessible paper requires
    & 25\%
    & 43\%
    \\ \hline
        arXiv requires submitting \\
        TeX, even if you have created \\
        an accessible PDF
    & 25\%
    & 25\%
    \\ \hline
        Lack of accessibility mandates
    & 25\%
    & 11\%
    \\ \hline
        Conference or journal \\
        guidelines are not accessible
    & 0
    & 11\%
    \\ \hline
        Not knowing how to make \\
        TeX accessible
    & 0
    & 25\%
    \\ \hline
        No access to authoring \\
        tools to create accessible \\
        content
    & 0
    & 11\%
    \\ \hline
    \end{tabular}
\end{table}

For users who do not use assistive technology the top reason was not knowing what an accessible paper requires, while for users of assistive technology it was that they consider their papers to be accessible already.

\rule{480pt}{.5pt}
\textsc{Preferred format}

\textbf{Summary:} Respondents who use assistive technology preferred HTML, while those who didn't preferred PDF. Use of specific types of assistive technology, such as screen magnification and color and contrast remediation, correlated with a strong preference for HTML.

We asked survey respondents what their preferred format is for reading papers. Interestingly, the survey responses around format contradict the results from our interviews, where HTML was very strongly indicated as the preferred format for accessibility. We conjecture that this difference in result is due to several factors:

\begin{enumerate}
    \item Our interview participants include a high percentage of accessibility and standards professionals, while the survey respondents are predominantly researchers who are less familiar with accessibility and the pros and cons of different formats.
    \item It is difficult to predict future behaviour, even our own. The dominance of the PDF format in scientific publishing—and its resultant workflows—makes it difficult to imagine alternatives.
    \item Many researchers like PDF. Based on user feedback it is clear that new formats for publishing must appear alongside existing options so researchers can interact with papers in their preferred way.
\end{enumerate}

When asked what their preferred format would be \emph{if well formatted HTML was available}, 67\% still indicated PDF would be preferred; among assistive technology users a small majority of 55\% would prefer HTML:
\pagebreak
\begin{table}[!ht]
    \centering
    \begin{tabular}{lll}
    \hline
       Preferred format
    & Yes, use assistive technology
    & No, don’t use assistive technology
    \\ \hline
        HTML
    & 55\%
    & 29\%
    \\ \hline
        PDF
    & 45\%
    & 67\%
    \\ \hline
        TeX
    & 0\%
    & 4\%
    \\ \hline
    \end{tabular}
\end{table}

When respondents were asked which format would be \emph{most useful} to them, there was a strong split between those who use assistive technology, and those who do not:

\begin{table}[!ht]
    \centering
    \begin{tabular}{lll}
    \hline
       Useful format
    & Yes, use assistive technology
    & No, don’t use assistive technology
    \\ \hline
        Well formatted HTML
    & 45\%
    & 16\%
    \\ \hline
        Well formatted PDF
    & 45\%
    & 69\%
    \\ \hline
    \end{tabular}
\end{table}

Interestingly, respondents with specific barriers indicated a preference for HTML: those who need to adjust font size, color, or contrast; all barriers which indicate a vision impairment.

Also preferring HTML are professors who help students with translation or by describing images and charts. One professor described how HTML would help him: \emph{“I translate material into braille for one user, which is highly specialized, but starting from html is much better than pdf.”}

\rule{480pt}{.5pt}
\textsc{Additional suggestions}

Respondents were asked to select from a list of potential site improvements which they would find useful. The following two changes had the highest positive response rate:
\begin{enumerate}
    \item The ability to build a customizable arXiv feed (73\%), and
    \item A quick way to get to a paper’s conclusions and references (58\%).
\end{enumerate}

Other suggestions included:
\begin{itemize}
    \item Coverage of more scientific fields
    \item More use of AI and machine learning for categorization and discovery
    \item The option to enlarge fonts
    \item Improve TeX upload function
    \item Add dark mode, and
    \item To keep arXiv going:
\end{itemize}

\textbf{“I REALLY appreciate the arXiv service. I can't afford to subscribe to a plethora of physics journals. Hats off to the arXiv team!”}
\pagebreak

\subsection{Interviews}
To better understand the ecosystem arXiv is operating in, we interviewed a wide range of researchers from the larger arXiv community.

\subsubsection{Process}
We interviewed a total of 44 individuals. They include researchers with reading disabilities or other access barriers who use arXiv, professors who help students with disabilities, researchers whose focus is on various fields of accessibility, experts on standards for web content, and leaders in LaTeX, MathML, and other languages critical to the success of this project. Some participants fit into more than one category.

Our participants were recruited through our accessibility survey, two accessibility mailing lists, direct invitations, and word of mouth. The mode for interview length was 30 minutes with some going longer.

We conducted semi-structured interviews with individuals or small groups using video conferencing tools, and in two cases in person. An interview guide was developed and used as a loose guiding tool, but prioritizing the participants' conversational lead: we wanted to learn what they most wanted us to know. All participants were interviewed by the main interviewer, assisted by other members of arXiv staff and student assistants.

All personally identifying data relating to interviewees is omitted in this report in order to preserve anonymity.

\subsubsection{Demographics}
Our participants were diverse in terms of their career stage and included PhD students, professors, and researchers working in industry. Participants come from multiple fields of research including physics, math, statistics, computer science, legal, and regulatory. We interviewed a number of participants who serve on various W3C boards related to accessibility and web standards.

Participants were not asked to disclose their disability, but were asked to describe if and how their disability affected their access to research. In responding to this question participants disclosed the following: 7 participants disclosed blindness, 1 participant disclosed dyslexia, 1 participant disclosed ADHD, and 2 participants disclosed a movement disability.

\subsubsection{Analysis}
All interviews were transcribed by either the main interviewer, or by student assistants and then reviewed by the main interviewer. Transcripts were broken down into observations, then documented following Atomic Research principles by mapping each observation to a semantic layer to facilitate discovery.

All observations were anonymized prior to thematic analysis to protect the privacy of our interviewees. During analysis we looked for similarities, grouped them into themes, and compared to data from the survey and looked for disparities and correlations.

\subsubsection{Interview Results}

{\Large\textbf The Accessible User Journey}

To evaluate the user journey we sorted feedback into five primary user goals: Find Research, Read Research, Participate in Scholarship, Prepare my Document, and Submit:

\begin{figure}[htp]
    \centering
    \includegraphics[scale=0.5]{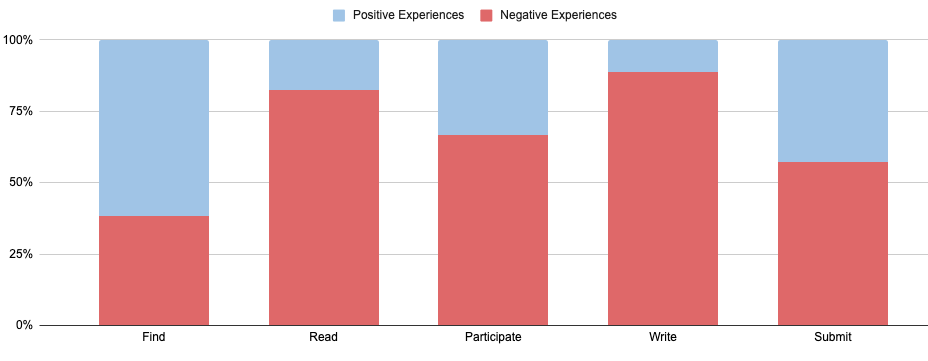}
    \caption{This graphic displays a stacked bar chart for each step of the user journey showing the percent of positive vs. negative experiences. Find research: 38\% negative, 62\% positive. Read papers: 83\% negative, 17\% positive. Participate in scholarship: 67\% negative, 33\% positive. Write papers: 89\% negative, 11\% positive. Submit papers: 57\% negative, 43\% positive.}
    \label{fig:stacked bar chart of each step of the user journey}
\end{figure}

 When analyzing each of these steps we asked: “when it comes to accessing research, was the experience of the participant positive or negative?” Most steps were dominated by negative experiences, with only Find Research being mostly positive. An accessible table display of the user journey, as well as the anonymized qualitative data behind it, is available in our \href{https://docs.google.com/spreadsheets/u/0/d/1Uvb-A1ePpYyWETxAizVI7ExHobCCSbuLczwxkFN6Ss4/edit}{User Journey Table}.

\emph{Read Research} elicited the highest number of comments, followed by \emph{Prepare my Document}:
\begin{figure}[htp]
    \centering
    \includegraphics[scale=0.4]{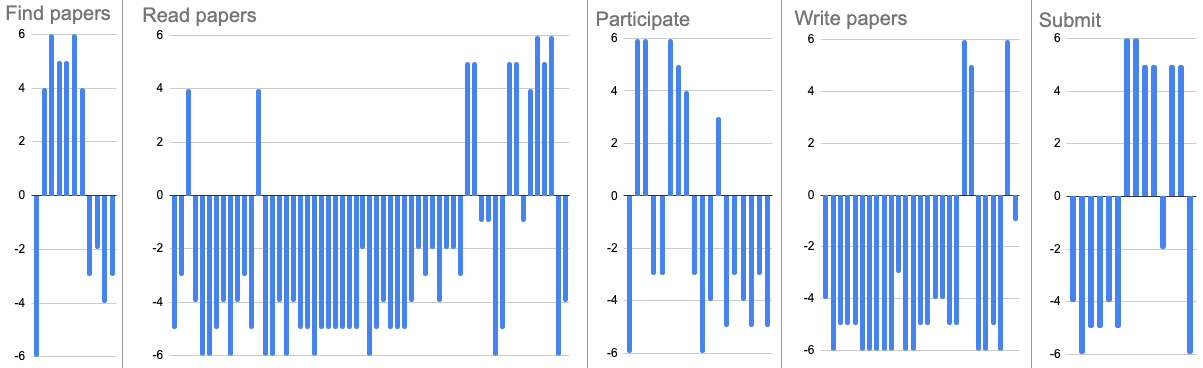}
    \caption{This graphic displays each of the five steps as an individual positive/negative bar chart. The number of bars at each step shows the number of experiences, while the height of each bar represent the impact score. Varying numbers of experiences were shared for each step, with Read eliciting the most by far. Find research: 12. Read papers: 57. Participate in scholarship: 17. Write papers: 27. Submit: 14.}
    \label{fig:positive/negative bar charts for each of the five steps in the user journey}
\end{figure}
\pagebreak

{\Large\textbf Themes}

Our thematic analysis of interview feedback identified 5 themes:
\begin{enumerate}
    \item The PDF format as a barrier.
    \item The benefits of HTML as a format.
    \item Skepticism on the potential for real change.
    \item arXiv has a role to play in improving the accessibility of research papers.
    \item HTML is just the beginning.
\end{enumerate}

\bigskip
\rule{480pt}{.5pt}
\textsc{Theme 1: researchers who rely on assistive technology point to PDF as the biggest stumbling block to accessible research.}

Participants who rely on assistive technology are particularly aware of—and frustrated with—its shortcomings:

\textbf{“Forget PDF!”}

\textbf{“I would prefer HTML over PDF.”}

\textbf{“You can make a pdf accessible but it is still a PDF (poor mobile use, no magnification).”}

\textbf{“PDF is not the best thing from an accessibility point of view.”}

\textbf{“[with PDF] I need to borrow someone and say ‘hey can you help me look at this data set.’”}

\textbf{“Screen reader does not do equations. Figures are a crapshoot.”}

\textbf{“Even accessible PDF, that just means they are tagged. But how do you read math in PDF?”}

\textbf{“I have walked away from PDF. Why are you wasting your time and mine?”}

For researchers who \emph{do not} rely on assistive technology, and for whom PDF already is sufficient, is is challenging to visualize accessing research in new ways that might break their current PDF-dependent workflow:

\textbf{“For scientific documents PDF is still excellent, in my opinion. I would not be that interested in HTML papers.”}

\textbf{“It is universal, and can keep things quite simple.”}

\textbf{“I generally prefer PDF… it’s useful to have it as a single file to add it to Zotero or wherever.”}

Some interviewees who initially dismissed HTML point out its benefits as they thought about it more:

\textbf{“I want the PDF so I can add it to Paper Pile… Although, [HTML] would make it easier to quote parts of a paper.”}

Making PDF accessible is challenging:

\textbf{“For PDF you need specialized people who understand the PDF standard. For tagging, you need tools from Acrobat and skilled people.”}

\bigskip
\rule{480pt}{.5pt}

\textsc{Theme 2: HTML mitigates a wide range of access issues, positively addressing many disabilities and improving the experience for most users.}

HTML has a significant edge for researchers with disabilities:

\textbf{“I prefer HTML versions. As an assistive tech user I find it much faster to navigate.”}

\textbf{“HTML also works for the deaf community. English is a second language to deaf readers… they would rather get ASL. There are tools for converting HTML to ASL.”}

\textbf{“What I hear from colleagues in astronomy who are blind or low vision is that HTML is the preferred delivery mechanism. It is the most accessible.”}

\textbf{“HTML gives you a lot more freedom. You can use software to change colors or typefaces.”}

\textbf{“I don’t see how an accessible PDF can be better than accessible HTML. I don’t see the upside to it.”}

\textbf{“With the HTML version, you can throw MathJax on your website as well, then be able to render that MathML and have it spoken correctly to AT users.”}

\textbf{“Screen readers are more efficient while working with HTML.”}

HTML offers benefits beyond disability access, such as for machine learning and mobile phones:

\textbf{“The parsability and adaptability of HTML is better than PDF.”}

\textbf{“With HTML, [researchers] can adjust the content to their needs. I'm all for HTML.”}

\textbf{“The PDF format is not the best suited to mobile phones.”}

\textbf{“Reading pdf articles on a smartphone is anyway difficult.”}

\textbf{“HTML5 is the right basis for longevity.”}

HTML also offers greater reduction of legal risks. Though not yet specified, the consensus of the legal scholars we spoke to was that legal mandates in the USA for user-uploaded content are inevitable and already underway:

\textbf{“Long term, user uploaded content will need to be accessible. But the standards aren't there yet.”}

\textbf{“Given how large your scope is it’s good that you are doing this now… you have such a massive volume of material it could be a headache.”}

\textbf{“arXiv is covered by the ADA. There is no pass for third-party submitted content.”}

\bigskip
\rule{480pt}{.5pt}

\textsc{Theme 3: There is deep skepticism, and even a sense of despair, around the likelihood of real change.}

Researchers with disabilities feel their advocacy has gone unheard:

\textbf{“I have waited 20 years for PDF to do anything with accessibility. That is half a working lifetime.”}

\textbf{“Progress is not moving anywhere on accessible papers.”}

\textbf{“For a long time now accessibility has been viewed as a 'nice to have.' Pushing accessibility up the agenda is the biggest thing that can happen.”}

Making all research accessible is an enormous challenge that extends beyond arXiv:

\textbf{“arXiv can’t do it without the involvement of ordinary people.”}

\textbf{“Encouraging authors to tag will only be successful if the main players apply the new standards together.”}

\textbf{“It's an upward struggle because of how the blind reader space works. Screen readers want users to use their tool and don't provide APIs.”}

\textbf{“Even with much bigger technological steps you will still need buy-in from authors to do this.”}

We repeatedly heard we should not let the perfect be the enemy of good. Some improvement in accessibility is much better than no improvement:

\textbf{“There is no silver bullet. No point in waiting or looking for one, when you have tools that will do 80\% of the work now.”}

\textbf{“70\% great and another 15\% good is not a terrible success rate.”}

\textbf{“Even a small stride is a great leap for the people it helps. Don't be discouraged.”}

Participants called out the need for more authoring tools and education if this effort is to succeed:

\textbf{“I think one of the biggest problems that we have is the paucity of tools that make it easy to make HTML.”}

\textbf{“The biggest problems [in making research papers accessible] is communicating the importance of this to authors. I hope awareness will grow over time.”}

\textbf{“There needs to be more [accessibility] guidance, especially if you want authors to do some of the work.”}

\bigskip
\rule{480pt}{.5pt}

\textsc{Theme 4: We heard from our community that arXiv can have a powerful and immediate impact.}

We are part of an ecosystem, and we have a role to play in making research accessible:

\textbf{“arXiv is well set up [to impact accessibility] in a way that a single publisher isn't.”}

\textbf{“Your job at arXiv is just to put out proper HTML.”}

\textbf{“In a dreamworld every technical document will just work [with screen readers]. In a smaller dream world, it just works on arXiv.”}

\textbf{“My area in physics is stuck in its ways. Not many people want to go against the norm. And arXiv sets the norm.”}

Because arXiv has direct control over the submission pipeline, it opens up opportunities:

\textbf{“The good news is that the potential for significant impact is just right there. The fact that you are interested in this is very exciting to me. We could shift how we author papers in a way that doesn't add burden but makes life easier. It is a value-add for everyone in authoring and output.”}

\textbf{“If someone is in control of the pipeline there is a chance accessibility will get built in.”}

\textbf{“There is no polished product you can buy somewhere but all the pieces are coming together. And if we can show the user what they need to do on their end, this could work very beautifully. ArXiv has the content, and you are compiling your own content, and you have the traffic.”}

\textbf{“Your push on the pipeline-based tool is the right way because you cannot scale up on it in any other way for it to be automated.”}

\bigskip
\rule{480pt}{.5pt}

\textsc{Theme 5: HTML is just one step towards a greater vision of accessibility.}

Research is consumed in different ways. The ultimate goal is flexibility and giving readers the ability to access the content in the way that works best for them:

\textbf{“In a perfect world I could be given the source files and reproduce and be in total control of the method I choose to consume.”}

\textbf{“I have a right to the information that the author was trying to communicate… There is an ethos in the blind community in [Country] that everyone consumes PDF, so we have the right to consume it as well. I believe I have the right to consume the content.”}

\textbf{“It really helps when people include the source.”}

\textbf{“I have written an add-on that can sonify a series... higher pitch for higher values and so on.”}

\textbf{“The ideal is that I get to choose how much information I skip over in the same way a sighted person does.”}

Current accessibility standards are only a starting point:

\textbf{“Accessibility has limitations; it is the bare minimum ground in its standards and is too legally driven. But the work of real access, making things work for people with disabilities, requires more.”}

Many interviewees with disabilities pointed out that a paper is just a portal to the knowledge behind it. They don’t necessarily need access to the paper if they can absorb the research itself:

\textbf{“Anything like tables are super helpful for me. Source data, or code, is much easier to understand. Then you don’t need to read the paper. If it is well written code then I would prefer that to reading the paper.”}

\textbf{“I would prefer if there was raw data available in an excel spreadsheet or similar. Then I can find how to make sense of the data. I would compile the statistics.”}

\section{arXiv's Plan for accessible research papers}
HTML is an even older and more established standard than PDF. In fact, HTML was invented to facilitate the sharing of scientific knowledge (CERN). It has well defined criteria for achieving accessibility on the web and does not require proprietary tools to author or consume. HTML also provides a better foundation for machine readability, and can help usher in the next generation of tools that will help us all find and access research more efficiently.

Of course, HTML is not automatically accessible. When we refer to HTML in our accessibility plan we mean well-formatted, semantic HTML with necessary ARIA tags. There will be some limitations to what we can do based on the richness of the original uploaded TeX, but our plan is to achieve the most accessible HTML possible within those constraints.

PDF can theoretically be tagged for accessibility, too. But as presented in this paper, the reality is that only about 2.4\% of PDFs are accessible and there are significant barriers to improving that number. One promising effort is that the LaTeX core working group are addressing accessibility now, and we plan on incorporating their work as soon as it is available so that we can also provide more accessible PDF documents on arXiv.
\pagebreak

Based on provided feedback, we have rated how well we expect the PDFs that arXiv generates now vs the well-formatted HTML that our plan will provide to see how they score on a number of criteria in the table below.

\textbf{arXiv assessment of HTML and PDF}

\small \emph{Scale: 0=non-functional, 1=OK, 2=good, 3=excellent.}

\begin{table}[!ht]
    \centering
    \begin{tabular}{lll}
    \hline
        Criteria
    & PDF
    & Well-formatted HTML
    \\ \hline
        Screen reader legible text
    & 2
    & 3
    \\ \hline
        Screen reader legible math
    & 1
    & 3
    \\ \hline
        Screen reader legible charts
    & 0
    & 0-1
    \\ \hline
        Screen reader legible images
    & 0
    & 0-1
    \\ \hline
        Screen magnifier compatibility
    & 0
    & 3
    \\ \hline
        Colors and contrast adjustment
    & 0
    & 3
    \\ \hline
        Mobile friendly
    & 0
    & 3
    \\ \hline
        Machine readability
    & 1
    & 3
    \\ \hline
        Portability
    & 3
    & 2
    \\ \hline
        Archival nature
    & 3
    & 3
    \\ \hline
        Ability to make accessible
    & 1
    & 2
    \\ \hline
        Established use in academia
    & 3
    & 0
    \\ \hline
        Open source
    & 1
    & 3
    \\ \hline
        Adjustable content*
    & 0
    & 3
    \\ \hline
        Legal risk mitigation
    & 1
    & 3
    \\ \hline
        Total score
    & PDF: 16
    & HTML: 35
    \\ \hline
    \end{tabular}
\end{table}

\emph{*examples: to meet publishers' requests to hide author names while the paper is under double blind review, or promoting best practices by displaying the license the author chose during submission on the work itself}

To offer the flexibility of well formatted HTML downstream requires, ironically, restrictions upstream during content creation. Well structured, parseable content that follows established standards must be either provided or generated during submission.

And this is the difficulty. 90\% of arXiv submissions are provided as TeX (mainly LaTeX), and converting to HTML is not easy due to its extensibility.

\quotetext{On the one hand it's great that LaTeX is so extensible. On the other hand it is such a pain that it is so extensible.}
\quoteattr{—Frank Mittelbach, Head of Development, LaTeX group}

Incorporating the conversion into arXiv’s submission process will mean substantial changes to the pipeline behind the scenes, not visible to the author or affecting their submission experience. However we will need to lean on author engagement in two ways: to add alternative text for images and other content that can’t be parsed, and to view and approve their HTML output before submitting, just as they do their PDF output today. arXiv will then make this content available directly on the website alongside the PDF and TeX source.

Because the relationship between authors and the arXiv platform is direct—with no third party typesetting the documents before publishing—we have a tremendous opportunity to make small changes in the submission pipeline that have profound results on accessibility.

\quotetext{arXiv has a closeness to the practitioner that is exciting for accessibility. A lot of remediations require a human touch.}
\quoteattr{—Frank Elavsky, Data Visualization and Accessibility expert}

\section*{Conclusions}
The level of accessibility of research papers is low, and we cannot claim to have achieved truly open science while those with disabilities are barred from equivalent access. Based on our user research, the step our community wants arXiv to take is clear: offer well formatted, accessible HTML alongside existing sources.

Adding HTML will allow all researchers to experience its benefits, try new workflows, and adjust how papers are authored over time. It will support existing and emerging assistive technologies that work most efficiently with HTML, and normalize the format across more fields.

It is not an easy goal, but it is an achievable one. And because of arXiv’s reach across many fields and control over the submission pipeline we are positioned to leverage HTML in an impactful way.

\section*{Acknowledgements}
This material is based upon work supported by the National Science Foundation under Award No. OAC-2311521 and by NASA under award No. 20-OSTFL20-0053. Any opinions, findings and conclusions or recommendations expressed in this material are those of the author(s) and do not necessarily reflect the views of the National Science Foundation or of NASA.

\section*{References}
Adobe. “The case for reimagining reading.” Adobe Blog, 5 December 2018, \sloppy
\url{https://blog.adobe.com/en/fpost/2020/reimagining-reading-infographic}.

Bingham, Jeffrey P., et al. “An Uninteresting Tour Through Why Our Research Papers Aren’t Accessible.” ACM CHI, 2016. \sloppy
\url{http://dx.doi.org/10.1145/2851581.2892588}.

Centers for Disease Control and Prevention. “Burden of Vision Loss.” Centers for Disease Control and Prevention, 2020, \sloppy
\url{https://www.cdc.gov/visionhealth/risk/burden.htm}.

CERN. “The birth of the Web.” CERN, \sloppy
\url{https://home.cern/science/computing/birth-web}.

Code.org Advocacy Coalition, et al. “2022 State of Computer Science Education.” Code.org Advocacy Coalition, 2022, \sloppy
\url{https://advocacy.code.org/2022\_state\_of\_cs.pdf}.

Elavsky, Frank, et al. “How accessible is my visualization? Evaluating visualization accessibility with Chartability.” Computer Graphics Forum, vol. 41, no. 3, 2022, \sloppy
\url{https://www.frank.computer/chartability/}.

Hrastinski, Iva, and Ronnie B. Wilbur. “Academic Achievement of Deaf and Hard-of-Hearing Students in an ASL/English Bilingual Program.” Journal of Deaf Studies and Deaf Education, vol. 21, no. 2, 2016. \sloppy
\url{https://www.jstor.org/stable/26172441}.

Mankoff, Jennifer, et al. “2019 Access SIGCHI report.” ACM SIGACCESS Accessibility and Computing, no. 126, 2020. \sloppy
\url{https://doi.org/10.1145/3386280.3386287}.

National Center for Education Statistics. “Children 3 to 21 years old served under Individuals with Disabilities Education Act (IDEA), Part B, by age group and sex, race/ethnicity, and type of disability: 2020-21.” National Center for Education Statistics, 2021, \sloppy
\url{https://nces.ed.gov/programs/digest/d21/tables/dt21\_204.50.asp}.

Policy and Global Affairs, et al. Automated Research Workflows for Accelerated Discovery: Closing the Knowledge Discovery Loop. National Academies Press, 2022, \sloppy
\url{https://nap.nationalacademies.org/catalog/26532/automated-research-workflows-for-accelerated-discovery-closing-the-knowledge-discovery}. Accessed 14 October 2022.

United Nations. “Factsheet on Persons with Disabilities.” United Nations, Department of Economic and Social Affairs, \sloppy
\url{https://www.un.org/development/desa/disabilities/resources/factsheet-on-persons-with-disabilities.html}. Accessed 28 September 2022.

Vegas, Emiliana, and Brian Fowler. “What do we know about the expansion of K-12 computer science education?” Brookings Institution, 4 August 2020, \sloppy
\url{https://www.brookings.edu/research/what-do-we-know-about-the-expansion-of-k-12-computer-science-education/}.

Wang, Lucy, et al. “Improving the accessibility of scientific documents.” arXiv, 2021. arxiv.org, \sloppy
\url{https://arxiv.org/pdf/2105.00076}.

“Website Accessibility Regulations On The Horizon: DOJ To Start Title II Rulemaking For State and Local Governments Next Year.” ADA Title III, Seyfarth, 29 July 2022, \sloppy
\url{https://www.adatitleiii.com/2022/07/website-accessibility-regulations-on-the-horizon-doj-to-start-title-ii-rulemaking-for-state-and-local-governments-next-year/}.

Wentz, Brian, et al. “A Socio-Legal Framework for Improving the Accessibility of Research Articles for People With Disabilities.” Journal of Business \& Technology Law, vol. 16, no. 223, 2021. DigitalCommons@UM Carey Law, \sloppy
\url{https://digitalcommons.law.umaryland.edu/cgi/viewcontent.cgi?article=1333\&context=jbtl}

World Health Organization. World Report on Vision. 2019, \sloppy
\url{https://www.who.int/publications-detail-redirect/9789241516570}.

Yale Center for Dyslexia and Creativity. “Dyslexia FAQ.” Yale Center for Dyslexia, \sloppy
\url{https://dyslexia.yale.edu/dyslexia/dyslexia-faq/}.

\end{document}